# Reflections on the Fifth International Conference on Women in Physics

Chandralekha Singh, Department of Physics and Astronomy, University of Pittsburgh

**Abstract.** This article describes reflections on the Fifth International Conference on Women in Physics which was a conference attended by 215 female physicists and a few male physicists from 49 different countries. The article focuses on the barriers that women face in their professional advancement in physics and the extent to which the situation is different in various countries.

Imagine women physicists from every country on the planet! I made it my mission to meet and talk to as many of them as possible, and take note of their experiences. The countries represented were big and small. I had the opportunity to interact with women physicists not only from the Western countries but also from Burkina Faso, Estonia, Algeria, Tunisia, Iran, Egypt, Zambia, Tanzania, Ghana, Pakistan, Nepal, India, Lithuania, Armenia, Sudan, Japan, China, Uganda, Belarus, Cameroon, Ethiopia, Nigeria, Uruguay, Ecuador, Mexico, El Salvador, Honduras, and Philippines.

In August 2014, I attended the 5th International Conference on Women in Physics (ICWIP2014) in Waterloo Canada as a part of the US delegation. The conference was attended by approximately 215 female physicists and a few male physicists from 49 different countries. There were research talks, panels, workshops, breakout sessions and posters on issues related to women in physics.

Barriers to the advancement of women in physics that were listed in the country reports included societal biases affecting women and accumulating over time from an early age, unconscious gender bias, stereotype threat, family responsibilities, unfriendly and unsupportive environment in physics departments, lack of mentoring, lack of a critical number of women in physics and role models, physics being a competitive field rather than a collaborative field, a perception that physicists do not try to explain to students how physics helps humankind, men in physics in some countries acting "macho" and women physicists feeling marginalized. And, in some countries, e.g., in the Sudan, issues that negatively impact and limit women in physics include religion, economics and politics.

There were workshops in which we learned what social scientists have ascertained about how girls are influenced as they grow up with regard to pursuing science and mathematics. In the workshop, "Equity and education: Examining gendered stigma in science", we learned that, while most girls are interested in science and math when they are in early grades, in countries like the US, many tend to step away, often because they unwittingly conform to societal gender stereotypes. Women in some countries like the US are often victims of gender stereotypes from very early on, and some women are impacted so much that they even start questioning their own ability to ever be equal to or better than men in STEM fields such as physics. Societal biases related to women not being smart enough to pursue careers in male dominated STEM fields can impact women's beliefs about their own capability and negatively influence whether women pursue STEM majors and how they perform in STEM courses.

In some countries such as the US, when women don't succeed in a science course, people often attribute it to their poor abilities; but when men do not succeed, people often attribute it to their lack of effort or poor teaching but not to their lack of ability. This dichotomy has a negative impact on whether women who have failed once ("failing" could even be obtaining a B or a C grade in a course for an otherwise high achieving woman) would want to pursue those subjects in the future. Due to these gender stereotypes, many women in male-dominated fields assume that small setbacks, e.g., getting one B or C grade in a physics course, are indicative of their lack of aptitude for physics. They are more likely to interpret such setbacks to imply that they are not cut out to pursue a physics related degree and so they lose confidence. If women underperform, they are often likely to blame themselves and feel that they do not have the talent necessary for excelling in the subject in which their male counterparts seem to have an edge. In several studies, if students did not perform well in a test and were told that learning is about effort, they tried harder and did better later but if they were told that learning is tied to innate ability, they did not try harder after they performed poorly.

The stereotype threat, e.g., directly or indirectly being reminded of the stereotype that physics is a field for brilliant men and women do not belong in physics can exacerbate the situation. Women become victims of stereotype threat when their performance is negatively impacted by their anxiety and negative perception about the group - "women"- to which they belong. For example, just asking women to write their gender on the test sheet before they take a test can act as a stereotype threat and can lead to women performing worse than they otherwise would if they were not told to write their gender. Writing her gender can act as a stereotype threat because women are already aware of the societal stereotype that women are not supposed to do as well as men in a math and science.

Such a threat often undermines a woman's ability to score high on tests or other standard measures of academic achievement. Research in some Western countries such as the US suggests that people often perform much below their level when they conform to a stereotype under the threat.

Based upon presentations and discussions at the conference, discrimination women physicists face in the work place is overt in some countries and in some cases subtle. The differences are often entangled with and caused by how each culture views women. Women in physics in many countries are still often made to feel that they have chosen a wrong career path. Their progress and success is overlooked. Their opinions are often dismissed even if they are worthy of further discussions and future implementations based upon those ideas. Women physicists from many countries in Africa, Asia and South America reported that they even have to justify why they chose physics despite being a woman because of the macho culture and societal norms.

The "leaky pipeline" prevents women physicists in all countries from reaching the highest levels of our profession. The amount of leakage and at what stage it occurs varies significantly from country to country. In the US, women's participation in physics decreases precipitously from high school to college level and then again in the top leadership positions in physics. However, unlike the situation in many other countries, in the US, in the last decade, there is no leak from the undergraduate to graduate to assistant professor level in physics --the percentage of women at each of these levels has hovered around 20%.

Regardless of the country, the common theme at the conference was that women are highly underrepresented in leadership positions and decision-making roles. The overall proportion of female researchers in Estonia is over 40% and exceeds the European average but the gender imbalance in the researcher population increases with age. Women physicists from some Asian countries, e.g., China, noted that everything was fine up to graduate school and there was no significant barrier for women in physics until they obtained their Ph.D. After the Ph.D., there is a perception that women do not have the ability to be good physics professors, researchers or scientific leaders or that they should focus on their family rather than pursuing a high-profile career as a physicist. The glass ceiling was cited as a major factor why women fail to reach the top in physics across the world.

Igle Gledhill (South Africa) explored various aspects of our working environment, including intangible ones. Making the distinction between schemas and stereotypes (negative schemas), she introduced the dichotomous pair of schemas: Men are associated with agentic schema (proactive, independent, assertive) and women are associated with nurturative schema (nurturing, helpful, cooperative). She noted that workplaces tend to be agentic and being in the minority in a workplace increases the probability of encountering contradictory schemas. In order to recruit and retain more women in physics and improve the climate for women physicists, she emphasized the need to develop an awareness of our own schemas and challenge them. She noted that it is useful to be aware that prejudice rises during times of threat and, once it is in place, it lingers much longer than the threat does, and is hard to shake off. She also noted that research has shown that a stranger introduced to an environment will assume that a person with advantage is inherently better, whether that is the case or not. Once advantage and disadvantage have been established, the gap tends to widen which can hurt women physicists since they are unlikely to be in an advantageous position.

The other reason discussed for why women fail to rise and sometimes quit midway in physics is a sense of isolation. Women from many countries complained of lack of cooperation and even a condescending attitude displayed by many colleagues.

Casey Tesfaye (American Institute of Physics) presented the results of the International Survey of Physicists analyzed by regions and restricted to 12 countries. In nine of the analyzed countries, women had fewer opportunities than men and in a different nine country subset they had fewer resources than men. Regarding career progress, women with children progressed more slowly than men in eight of the analyzed countries.

There were several noteworthy differences between the experiences of women physicists in Western versus non -Western countries. The studies from North America suggest that family is not always (although it sometimes is) the biggest impediment to the professional advancement of women scientists; instead, the largest hurdles are the access to resources and implicit bias. However, some women physicists from non-Western countries noted that these findings do not apply to them. In many countries, even today, women are told from a very early age that they should only dream about getting married and having a husband and children, and so most grow up not having high professional aspirations and if they do, they encounter explicit gender bias while navigating family and professional aspirations.

While women in physics in Western countries are striving to be treated as equals without implicit biases and better opportunities for work life balance, in many countries in Africa, South America and Asia, the biases against female physicists are not subtle at all. Women who aspire professionally not only have difficulty fitting in at work; they may also struggle equally hard to win the support of their family members.

Women physicists, especially, from some developing countries, noted that taking an interest in physics is also perceived to diminish their feminine attributes. In fact, even in the US, the stereotyped portrayal of female scientists by popular media (e.g., the TV show "The Big Bang Theory") which make them look unattractive does not help in encouraging more young girls to pursue physics. Eileen Pollack, who wrote a New York Times opinion piece (October 13, 2013) about why there are so few women in science, attended this ICWIP2014 conference as a panelist and raised the point that the paucity of women going into physics is exacerbated by the stereotyped portrayal of female scientists.

Women physicists from Iran noted that more than 60% of BS and MS students, 47% of PhD students but only 18% of faculties in the physics departments are currently women. These high percentages of female physics students are partly because men in Iran are often more interested in engineering since the career prospects are better. Women from Egypt noted that the reason many women do not take comparable jobs to men even after obtaining their Ph.D. is that many want to be closer to home in order to take care of their families and have lower aspirations professionally in order to balance work and family.

What was clear is that in many of these non-Western countries the women physicists have an even greater difficulty balancing family and work. Not only are they responsible for everything at home, in addition, childcare and flexible work hour options are much less common in these countries. Some women physicists from those countries seemed resigned to the fact that they are unlikely to get an opportunity to pursue a career in physics which is as highly rewarding as the one afforded to their male counterparts because they have to find a job closer to home in order to balance work and life. In some of these countries, efforts to provide opportunities to balance work and family and counter biases that exacerbate the difficulties are impossible to even dream of at this time although they may be challenging in Western countries as well.

Even in western countries female physicists face a variety of challenges. The German contingent discussed data suggesting that female physicists' professional competence and accomplishments are less appreciated and that parenthood affects their education and career distinctly stronger than their male counterparts. Physicists from Finland (where the first female professor of physics was hired in 2004 at the University of Helsinki) noted that cultural reasons were central for understanding the gendered career segregation processes. For example, they noted that many major decisions are made in men only "saunas" which automatically excludes women physicists.

While all ICWIP2014 participants felt that early interventions at the K-12 level were important for the advancement of women in physics, there were large variations in the amount and effectiveness of these interventions implemented so far. Also, the societal stereotypes about who belongs in physics and can excel in physics are significantly disadvantaging women's advancement in physics. The physics culture must be fixed to make sure that women have the opportunities to excel in physics.

We discussed that many women often set high standards for whatever they do and if there are any lapses due to lack of support and physics culture, they may lose confidence in themselves and feel they have failed. What these women desperately need is more mentoring, encouragement, recognition, guidance and support from parents, teachers, counselors, role models and the entire physics community in order to persist and excel.

Cathy Foley (Australia) made a presentation, titled "Can you have it all? Making it work for you" and noted that if unconscious societal and workplace biases are removed and climate and support for women physicists at workplace improves, women can succeed both professionally and at home. In another session, Prajval Shastri (India) made a presentation titled "Interventions towards gender equity in physics: seeding or hindering cultural change?" in which she discussed why certain policies for advancement of women in physics are better than others. In the discussions, consequences of Germany's generous maternity leave policy emerged as an example supporting replacement of maternity leave by family leave. Finland emerged as a counter-example in that the generous gender-neutral parental leave policy is used by less than 10% of fathers. Beth Cunningham, AAPT Executive Officer, asked participants to engage with the issue of how physics societies can make a difference in the success of women in physics (WIP). She introduced the activities of AAPT and APS related to childcare grants, committees on WIP, workshops on WIP, and inventory of women in leadership roles, awards, plenary speakers, and editorial boards.

Diversity enhances excellence in physics and other STEM fields. This is especially important and urgent in an era in which the society has become increasingly dependent on science and engineering innovations at every walk of life. Conference participants agreed that we need to change the physics culture and provide appropriate support to everybody regardless of who they are from a very young age to cultivate their interest in these disciplines so that they feel like they belong in physics and can live up to their potential.

It is encouraging that in many countries there is more awareness in the science and engineering departments in which women are underrepresented, that there may be implicit and explicit biases and stereotypes that significantly impact the underrepresentation of women and more effort should be devoted to change the culture and recruit, support and retain talented women to ensure that everybody has an opportunity to contribute to the vitality of these disciplines.

At this conference, there were amazing bonds created. The experience was both uplifting and humbling at the same time. I shared many challenges that I have faced. I also felt that the experiences of women physicists in developed countries are challenging but not nearly as difficult as what some of these women physicists in developing countries routinely endure. I think the US can take a leadership role and lead the way with policies and practices that support women physicists, and forge a future by creating an equitable and inclusive learning environment. This type of leadership will also enhance excellence in physics. You can hear interesting and inspiring stories from some of the participants in the conference by watching a 14-minute video: https://www.youtube.com/watch?v=ofE-mJFJR5w

Note: Since I am posting this paper on arxiv for wider dissemination in 2022, I recommend that readers take a look at the following references [1-31] on related issues.

**Acknowledgement**

I am very grateful to the National Science Foundation for support.